\documentclass[fleqn,12pt]{article}
\usepackage{graphics,epsfig,here,cite}

\def\beq   {\begin{equation}}
\def\eeq   {\end{equation}}
\def\beqd  {\begin{displaymath}}
\def\eeqd  {\end{displaymath}}
\def\beqaa {\begin{eqnarray}}
\def\eeqaa {\end{eqnarray}}

\def\ti  {\tilde}

\def\sz{\ifmmode{\tilde{\chi}^0} \else{$\tilde{\chi}^0$} \fi}
\def\sw{\ifmmode{\tilde{\chi}} \else{$\tilde{\chi}$} \fi}

\newcommand{\be}[1]{\begin{equation} \label{(#1)}}
\newcommand{\ee}{\end{equation}}
\newcommand{\baq}[1]{\begin{eqnarray} \label{(#1)}}
\newcommand{\eaq}{\end{eqnarray}}
\newcommand{\rf}[1]{(\ref{(#1)})}
\newcommand{\ba}{\begin{array}}
\newcommand{\ea}{\end{array}}

\begin{document}
\pagestyle{empty}

\vspace*{-1cm} 
\begin{flushright}
  UWThPh-2007-11 
\end{flushright}

\vspace*{1.4cm}

\begin{center}

{\Large {\bf Effects of Lepton Flavour Violation
on Chargino Production at the Linear Collider 
}}\\

\vspace{2cm}

{\large 
Karl~Hohenwarter-Sodek and Thomas~Kernreiter}

\vspace{2cm}

{\it Faculty of Physics, University of Vienna, \\
Boltzmanngasse 5, A-1090 Wien, Austria}

\end{center}

\vspace{2cm}

\begin{abstract} 

We study the effects of lepton flavour violation (LFV)
on the production processes $e^+e^-\to\ti\chi^+_i\ti\chi^-_j$
at a linear collider with longitudinal $e^+$ and $e^-$ beam polarizations.
In the case of LFV the sneutrino mass eigenstates have 
no definite flavour, therefore, in the $t-$channel more than one sneutrino
mass eigenstate can contribute to the chargino production cross sections. 
Our framework is the Minimal Supersymmetric
Standard Model (MSSM) including LFV terms. 
We show that in spite of the restrictions on the LFV parameters
due to the current limits on rare lepton decays, the cross section
$\sigma(e^+e^-\to\ti\chi^+_1\ti\chi^-_1)$ can change by a factor 
of 2 or more when varying the LFV mixing angles.
We point out that even if the present bound on BR($\tau^-\to e^-\gamma$) 
improves by a factor of thousand the influence of LFV on the
chargino production cross section can be significant.
These results could have an important impact on the strategies for
determining the underlying model parameters at the linear collider.

\end{abstract}

\newpage
\pagestyle{plain}


\section{Introduction}

The Minimal Supersymmetric Standard Model (MSSM) \cite{mssm} includes
the spin--1/2 partners of the $W^\pm$ bosons and the charged
Higgs bosons $H^\pm$. These states mix and form the
charginos $\ti\chi^\pm_k$, $k=1,2$, as the mass eigenstates. 
The charginos are of particular interest, as they will
presumably be among the lightest supersymmetric (SUSY) particles. 
Therefore the study of chargino production
\be{eq:prodchar}
e^+e^-\to\ti\chi^+_i\ti\chi^-_j ~,\qquad i,j=1,2~,
\ee
will play an important role at the International Linear Collider (ILC).
This process has been studied extensively in the literature, see e.g. 
\cite{char1,char2,char3,char4,char5,char6,char7}.
Procedures have been developed to determine the underlying
parameters $\tan\beta$, $M_2$ and $|\mu|$, including the cosine of the
phase of $\mu$, $\cos\phi_\mu$, through
a measurement of a set of suitable observables in the processes
\rf{eq:prodchar}, where either various options for the beam polarizations 
are exploited \cite{char3,char4,char5} or spin correlations of the decaying charginos 
are studied \cite{char2,char5}. 
These studies assume that individual lepton flavour is conserved, which 
means that only one sneutrino ($\tilde\nu_e$) contributes to the 
processes \rf{eq:prodchar} via $t-$channel exchange.

In the present paper we study the influence of lepton flavour violation
(LFV) on the production cross sections  
$\sigma(e^+e^-\to\ti\chi^+_i\ti\chi^-_j)$ including longitudinal
beam polarizations. In general, the sizes of the SUSY LFV parameters are resticted 
as they give rise to LFV rare lepton decays at 1--loop level, which
have not been observed so far. From experimental searches upper bounds
on the branching ratios of these decays have been derived. 
For LFV muon decays the current limits are
BR$(\mu^-\to e^-\gamma) < 1.2 \cdot 10^{-11}$ \cite{Brooks:1999pu} and
BR$(\mu^- \to e^- e^+ e^-) < 1.0\cdot 10^{-12}$ \cite{Bellgardt:1987du} and for
the rate of $\mu^-- e^-$ conversion the best 
limit so far is $R_{\mu e}< 7.0\cdot 10^{-13}$ \cite{Bertl:2001fu}, with
$R_{\mu e}=\Gamma[\mu^-+N(Z,A)\to e^-+N(Z,A)]/
\Gamma[\mu^-+N(Z,A)\to \nu_\mu+N(Z-1,A)]$.
The sensitivities on LFV tau decays, on the other hand, are
smaller but have been improved substantially during the last years. 
The current limits for LFV tau decays are BR$(\tau^-\to e^-\gamma) < 1.1 \cdot 10^{-7}$ 
\cite{Aubert:2005wa}, BR$(\tau^-\to\mu^-\gamma) < 6.8 \cdot 10^{-8}$ \cite{Aubert:2005ye},
BR$(\tau^-\to e^-e^+e^-) < 2.0 \cdot 10^{-7}$ \cite{Aubert:2003pc} and
BR$(\tau^-\to\mu^-\mu^+\mu^-) < 1.9 \cdot 10^{-7}$ \cite{Aubert:2003pc,Yusa:2004gm}.

It is the aim of this paper to demonstrate that in spite of the
restrictions due to LFV rare lepton decays the 
production cross sections $\sigma(e^+e^-\to\ti\chi^+_i\ti\chi^-_j)$ 
can change significantly when LFV parameters appear in the sneutrino system.
In particular we focus on the experimental situation 
where only the lightest chargino state is kinematically accessible
at a center of mass energy of 500~GeV. As we will show, the cross section for 
$e^+e^-\to\ti\chi^+_1\ti\chi^-_1$ can change by a factor of 2 and more in 
the presence of LFV.
This can be the case even if the present bounds on 
LFV rare lepton decays improve by three orders of magnitude.
If LFV effects of this size occur, then the minimal sets 
\cite{char2,char3,char4,char5} of observables may not be 
sufficient to determine the parameters in the chargino sector and have to be 
extended appropriately in order to take a possibly sizeable effect
of LFV into account.

The paper is organized as follows:
In section \ref{sec:2} we give a short account of sneutrino mixing 
in the presence of LFV.
In section \ref{sec:3} we present the formulae
for the cross sections of \rf{eq:prodchar} in the case
of LFV where the sneutrino $t-$channel contribution
has to be modified as compared to the case of lepton flavour conservation. 
We carry out a numerical analysis of the influence
of LFV on the chargino production cross sections in
section \ref{sec:4}. Section \ref{sec:5} contains our conclusions.

\section{Sneutrino mixing\label{sec:2}}

The sneutrino mass matrix in the MSSM including lepton flavour violation,
in the basis $(\tilde\nu_{e},\tilde\nu_{\mu},\tilde\nu_{\tau})$, 
is given by
\begin{eqnarray}
M^2_{\tilde \nu,\alpha\beta} &=&  M^2_{L,\alpha\beta} 
+ \frac{1}{2}~m^2_Z~\cos2\beta~\delta_{\alpha\beta} \qquad (\alpha,\beta=1,2,3)~.
\label{eq:sneutrinomass}
\end{eqnarray}
The indices $\alpha,\beta,\gamma=1,2,3$ characterize the flavours 
$e,\mu,\tau$, respectively.
$M^2_{L}$ is the hermitean soft SUSY breaking mass matrix for
the left sleptons, $m_Z$ is the mass of the $Z$ boson 
and $\tan\beta=v_2/v_1$ is the ratio of the vacuum expectation 
values of the Higgs fields.
The physical mass eigenstates are given by 
\begin{eqnarray}
\tilde \nu_i = R^{\tilde \nu}_{i\alpha}~\tilde\nu_\alpha' \qquad
(i=1,2,3)~,
\label{eq:Mixing}
\end{eqnarray}
with $\tilde \nu_\alpha'=(\tilde\nu_e, \tilde\nu_\mu, \tilde\nu_\tau)$.
The mixing matrix and the physical mass eigenvalues are obtained 
by an unitary transformation  
$R^{\tilde\nu}M^2_{\tilde \nu}R^{\tilde\nu\dagger}=
{\rm diag}(m^2_{\tilde\nu_1},m^2_{\tilde\nu_2},m^2_{\tilde\nu_3})$,
where $m_{\tilde\nu_1} < m_{\tilde \nu_2} < m_{\tilde \nu_3}$.
Clearly, for $M_{L,\alpha\neq\beta}\neq 0$
the mass eigenstates, Eq.~(\ref{eq:Mixing}), are not flavour eigenstates. 

\section{Cross section\label{sec:3}}

\begin{figure}[t]
\hspace{1cm}
\begin{minipage}[t]{3.5cm}
\begin{center}
{\setlength{\unitlength}{1cm}
\begin{picture}(1.5,2.5)
\put(-2.,-1.1){\includegraphics{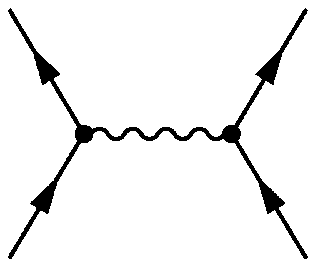}}
\put(-1.8,-1.6){$e^{-}(p_1)$}
\put(1,-1.6){$\tilde{\chi}^{-}_j(p_4)$}
\put(-1.8,1.5){$e^{+}(p_2)$}
\put(1,1.5){$\tilde{\chi}^{+}_i(p_3)$}
\put(-.3,.4){$\gamma$}
\end{picture}}
\end{center}
\end{minipage}
\hspace{2cm}
\vspace{.8cm}

\begin{minipage}[t]{3.5cm}
\begin{center}
{\setlength{\unitlength}{1cm}
\begin{picture}(-2.5,2.5)
\put(3.,2.3){\includegraphics{Feyn1.eps}}
\put(3.,1.7){$e^{-}(p_1)$}
\put(6.,4.9){$\tilde{\chi}^{+}_i(p_3)$}
\put(3.,4.8){$e^{+}(p_2)$}
\put(6.,1.8){$\tilde{\chi}^{-}_j(p_4)$}
\put(4.7,3.7){$Z$}
\end{picture}}
\end{center}
\end{minipage}
\hspace{2cm}
\vspace{.8cm}

\begin{minipage}[t]{3.5cm}
\begin{center}
{\setlength{\unitlength}{1cm}
\begin{picture}(2.5,2.5)
\put(11.8,5.4){\includegraphics{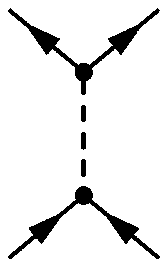}}
\put(11,8.2){$e^{+}(p_2)$}
\put(11,5.2){$e^{-}(p_1)$}
\put(13,8.2){$\tilde{\chi}^{+}_i(p_3)$}
\put(13,5.1){$\tilde{\chi}^{-}_j(p_4)$}
\put(12.1,6.6){$\tilde{\nu}_a$}
 \end{picture}}
\end{center}
\end{minipage}
\vspace{-4.5cm}
\caption{\label{bild1} 
Feynman diagrams for chargino production in $e^+e^-$-collisions.}
\end{figure}

The Feynman diagrams contributing to process \rf{eq:prodchar} 
are depicted in Fig.~\ref{bild1}. In the case of LFV the 
sneutrino contribution has to be modified, as now 
more than one sneutrino couples to the electron and positron 
(unless LFV arises solely due to the parameter $M^2_{L,23}$). 
The relevant part of the interaction Lagrangian 
which gives rise to the $t-$channel sneutrino contribution
is given by \cite{mssm,char7}
\be{eq:lagsneuchar}
\mathcal{L}_{\ell\tilde{\nu}\tilde{\chi}^+}=
-g~V^*_{j1}~R^{\tilde\nu*}_{a1}~\overline{\tilde{\chi}^{-}_j}~
P_L~e~\ti\nu_a^\dagger
-g~V_{j1}~R^{\tilde\nu}_{a1}~\bar{e}~P_R~\tilde{\chi}^{-}_j
~\ti\nu_a~,
\ee
where $P_{L,R}=1/2(1\mp\gamma_5)$, $g$ is the weak coupling constant 
and the unitary $2\times2$ mixing matrices $U$ and $V$ 
diagonalize the chargino mass matrix ${\mathcal M}_C$, 
$U^{\ast}{\mathcal M}_C V^{-1}=
{\rm diag}(m_{\chi_1},m_{\chi_2})$. 
In Eq.~\rf{eq:lagsneuchar} we have omitted terms that
are proportinal to the tiny electron Yukawa coupling.

The production cross section for the process \rf{eq:prodchar} is given by
\be{eq:crossectionprod}
{\rm d}\sigma=\frac{1}{2 (2 \pi)}
\frac{q}{s^{3/2}}~P~ {\rm d}\cos\theta~,
\ee
where $\sqrt{s}$ is the cms energy, $q$ is the momentum of 
the $\ti\chi^{\pm}$'s and $\theta$ is the scattering angle.
$P$ is the amplitude squared averaged and summed over 
the polarizations of the produced charginos.
We closely follow the notation of \cite{char7} where
$P$ is given by the terms 
\be{eq:P}
P=P(\gamma \gamma)+P(Z Z)+P(\gamma Z)+ 
\sum^3_{a=1}( P(\gamma \tilde{\nu}_a)+ P(Z\tilde{\nu}_a) )+
\sum^3_{a=1} \sum^3_{b=1} P(\tilde{\nu}_a\tilde{\nu}_b)~.
\ee
The terms that are modified in the presence of LFV
are the one involving the sneutrino exchange in the $t-$channel which read
\baq{eq:Pgs}
P(\gamma \tilde{\nu}_a) &=& c_L
\frac{g^4}{2}\sin^2\Theta_W 
{\rm Re}\{\Delta(\gamma)\Delta^*(\tilde\nu_a) 
|R^{\tilde\nu}_{a1}|^2 V^*_{i1}~V_{j1}~[2(p_1p_4)(p_2p_3)\nonumber\\
&&{}+m_{\chi_i} m_{\chi_j}(p_1p_2)] \delta_{ij}\}~,
\eaq
\baq{eq:PZs}
P(Z \tilde{\nu}_a) &=& c_L
\frac{g^4}{2}\tan^2\Theta_WL_e 
{\rm Re}\{\Delta(Z)\Delta^*(\tilde\nu_a)|R^{\tilde\nu}_{a1}|^2
V^*_{i1}V_{j1}[O'^R_{ij} m_{\chi_i} m_{\chi_j}(p_1p_2)\nonumber\\
&&{}+2 O'^L_{ij} (p_1p_4)(p_2p_3)]\}~,
\eaq
\be{eq:Pss}
P(\tilde{\nu}_a \tilde{\nu}_b)=c_L
\frac{g^4}{4} 
\Delta(\tilde\nu_a)\Delta^*(\tilde\nu_b)
|R^{\tilde\nu}_{a1}|^2|R^{\tilde\nu}_{b1}|^2
|V_{i1}|^2|V_{j1}|^2(p_1p_4)(p_2p_3)~,
\ee
with
\begin{equation}
O'^L_{ij}=-V_{i1}V^*_{j1}-\frac{1}{2}V_{i2}V^*_{j2}+
\delta_{ij}\sin^2\Theta_W~,
\end{equation}
\begin{equation}
O'^R_{ij}=-U^*_{i1}U_{j1}-\frac{1}{2}U^*_{i2}U_{j2}+
\delta_{ij}\sin^2\Theta_W~,
\end{equation}
\begin{equation}
L_e=-\frac{1}{2}+\sin^2\Theta_W~,
\end{equation}
with $\Theta_W$ being the Weinberg angle.
The propagators in Eqs.~\rf{eq:Pgs}--\rf{eq:Pss} are given by
$\Delta(\gamma)=i/s$, $\Delta(Z)= i/(s-m^2_Z), \Delta(\tilde{\nu}_a)=
i/(t-m^2_{\tilde{\nu}_a})$, with $s=(p_1+p_2)^2$, 
$t=(p_1-p_4)^2$. 
The assignment for the momentum vectors can be read off from Fig.~\ref{bild1}.
In Eqs.~\rf{eq:Pgs}--\rf{eq:Pss}, $c_L=(1-{\mathcal P}^-_L)(1+{\mathcal P}^+_L)$, 
where ${\mathcal P}^-_L$ (${\mathcal P}^+_L$) 
[$-1\leq {\mathcal P}^-_L, {\mathcal P}^+_L \leq 1$] denotes the degree
of the longitudinal polarization of $e^-$ ($e^+$). 
The remaining terms in Eq.~\rf{eq:P} can be found in \cite{char7}.
We note that in the limit of degenerate sneutrino masses an influence of LFV 
disappears, as we have $\Delta(\tilde{\nu}_1)=
\Delta(\tilde{\nu}_2)=\Delta(\tilde{\nu}_3)$ and
$\sum_{a=1}^3~|R^{\tilde\nu}_{a1}|^2=1$, see Eqs.~\rf{eq:Pgs}--\rf{eq:Pss}.
This is as expected, because in this case the three LFV mixing
angles in $R^{\tilde\nu}$ can be rotated away.

\section{Numerical analysis \label{sec:4}}

In the following we analyze numerically the influence of LFV
on the production cross section 
$\sigma(e^+e^-\rightarrow\tilde{\chi}^+_1\tilde{\chi}^-_1)$. 
We consider scenarios where only the pair production of the
lighter chargino states are kinematically accessible at a 
linear collider with a cms energy of $\sqrt{s}=500$~GeV.
We assume that a degree of beam polarization 
of $90\%$ for the electron beam and of $60\%$ for the
positron beam is feasible.
The LFV parameters on which
$\sigma(e^+e^-\rightarrow\tilde{\chi}^+_1\tilde{\chi}^-_1)$
sensitively depend are $M_{L,12}$ and $M_{L,13}$ and we discuss their influence 
separately. The LFV parameter $M_{L,23}$ has an influence
only if in addition $M_{L,12}$ and/or $M_{L,13}$ are non-vanishing, because
only the $R^{\tilde\nu}_{a1}$ elements appear in Eqs.~\rf{eq:Pgs}--\rf{eq:Pss}.

\subsection{$\tilde\nu_e$--$\tilde\nu_\tau$ mixing case}

We start the discussion assuming a non-vanishing $M^2_{L,13}$.
The size of $M^2_{L,13}$ is restricted by
the experimental upper bounds on the 
LFV processes $\tau^-\to e^-\gamma$ and $\tau^-\to e^- e^+ e^-$ to 
which it contributes at loop level. The formulae for the decay widths
of these reactions can be found in \cite{Hisano:1995cp}. For a complete 
1--loop calculation of the LFV leptonic three--body decays see \cite{Arganda:2005ji}.  
The decay width for the LFV leptonic two--body decays 
$\ell_j^-\to \ell_i^-\gamma$ ($\ell_j=\tau,\mu; \ell_i=\mu,e$), in the 
convention of \cite{Hisano:1995cp}, is given by
\begin{equation}
\Gamma(\ell_j^-\to
\ell_i^-\gamma)=\frac{\alpha}{4}~m^5_{\ell_j}~(|A^L|^2+|A^R|^2)~,
\label{eq:raredecays}
\end{equation}
with $\alpha=1/137$.
$A^L$ and $A^R$ are the left and right amplitudes, which
include the 1--loop contributions due to chargino--sneutrino exchange
and neutralino--slepton exchange. 
Furthermore, we require that the MSSM parameters have to 
respect the experimental limits of the anomalous magnetic
moments of the leptons, in particular that one of the muon, where
the difference between experiment and Standard Model (SM) prediction is 
$a^{\rm exp}_\mu-a^{\rm SM}_\mu=29\pm 9\cdot 10^{-10}$ \cite{Jegerlehner:2007xe}.
We impose that the SUSY contributions to $a_\mu$ 
must be positive and below $38 \cdot 10^{-10}$.  

The MSSM parameters on which the cross section 
$\sigma(e^+e^-\rightarrow\tilde{\chi}^+_1\tilde{\chi}^-_1)$ depends are the
parameters in the chargino sector $\mu$, $M_2$ and $\tan\beta$,
and the soft SUSY breaking mass parameters in the sneutrino sector $M_{L,11}$, $M_{L,22}$,
$M_{L,33}$ and $M_{L,13}$ ($M_{L,12}=M_{L,23}=0$ in this subsection).
In place of the SUSY parameters in the sneutrino sector we
treat the sneutrino masses $m_{\tilde\nu_1}$, $m_{\tilde\nu_2}$,
$m_{\tilde\nu_3}$ and the LFV mixing angle 
$\cos2\theta_{13}$ as our input parameters. This can be 
achieved by an inversion of the eigenvalue equations 
$R^{\tilde\nu}M^2_{\tilde \nu}R^{\tilde\nu\dagger}=
{\rm diag}(m^2_{\tilde\nu_1},m^2_{\tilde\nu_2},m^2_{\tilde\nu_3})$.

In addition to the MSSM paramters listed above the decay
widths of the rare lepton decays, Eq.~(\ref{eq:raredecays}), depend also 
on other MSSM parameters, which we fix throughout this study.
These are the soft SUSY breaking parameters in the charged slepton sector,
which we take as $M_{E,11}=700$~GeV, $M_{E,22}=800$~GeV, $M_{E,33}=900$~GeV, 
$M_{E,\alpha\neq\beta}=0$, $A_{\alpha\beta}=0$, $\alpha,\beta=1,2,3$,
(for the convention see e.g. \cite{Bartl:2005yy}), 
and the parameter $M_1$ of the neutralino sector, where
we assume the GUT inspired relation 
$|M_1|=(5/3)\tan^2\Theta_W~M_2$, with $M_1<0$ ($\phi_{M_1}=\pi$).

\begin{figure}[t]
\setlength{\unitlength}{1mm}
\begin{center}
\begin{picture}(150,50)
\put(-45,-145){\mbox{\epsfig{figure=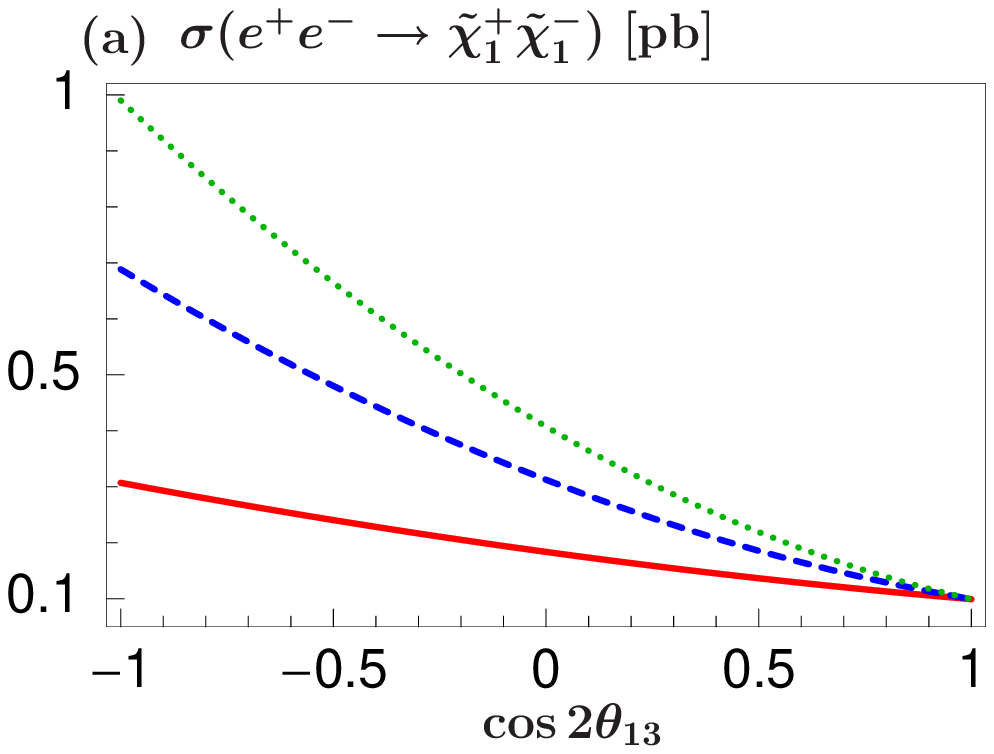,height=22.cm,width=14.4cm}}}
\put(30,-137){\mbox{\epsfig{figure=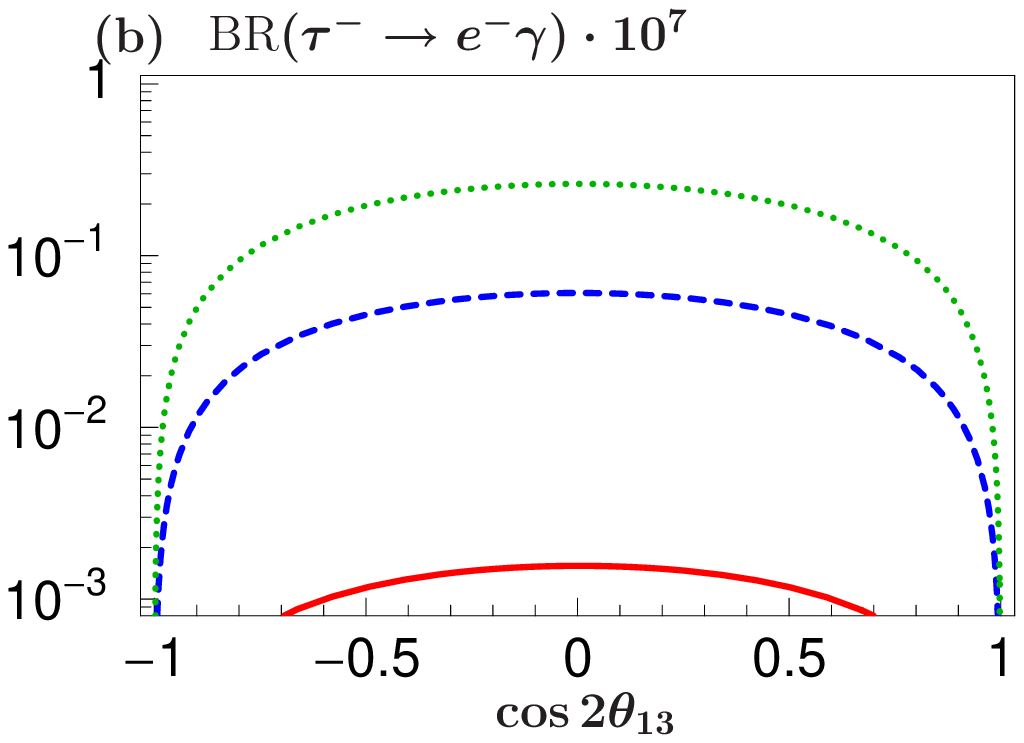,height=21.cm,width=14.4cm}}}
\end{picture}
\end{center}
\caption{(a) Cross section
$\sigma(e^+e^-\rightarrow\tilde{\chi}^+_1\tilde{\chi}^-_1)$ and
(b) branching ratio BR($\tau^-\to e^-\gamma$) as a function
of $\cos2\theta_{13}$. The three lines correspond to 
$m_{\tilde\nu_3}=400$~GeV (solid line),
600~GeV (dashed line) and 900~GeV (dotted line). 
The other parameters are as specified in the text.}
\label{fig:fig1}
\end{figure}

In Fig.~\ref{fig:fig1}a we show the $\cos2\theta_{13}$ dependence of the
cross section $\sigma(e^+e^-\rightarrow\tilde{\chi}^+_1\tilde{\chi}^-_1)$ 
for three values of $m_{\tilde\nu_3}=(400,600,900)$~GeV with $m_{\tilde\nu_1}=300$~GeV,
$m_{\tilde\nu_2}=350$~GeV, $\mu=1500$~GeV, $M_2=240$~GeV and $\tan\beta=5$.
The resulting chargino masses are $m_{\chi_1}=238$~GeV and $m_{\chi_2}=1505$~GeV.
The choice for the degree of beam polarizations is 
${\mathcal P}^-_L=-0.9$ and ${\mathcal P}^+_L=0.6$. Fig.~\ref{fig:fig1}b shows the 
corresponding dependence of the branching ratio 
BR($\tau^-\to e^-\gamma$) for the same parameters.
As can be seen in Fig.~\ref{fig:fig1}b, the LFV mixing angle
$\cos2\theta_{13}$ is not restricted and can have any value 
in the range $[-1,1]$. $\cos2\theta_{13}=-1,1$ are the cases 
where lepton flavour is conserved, while
for $\cos2\theta_{13}=0$ LFV is maximal, and the mass eigenstates 
$\tilde\nu_1$ and $\tilde\nu_3$ are mixtures containing an equal 
amount of $\tilde\nu_e$ and $\tilde\nu_\tau$.

Furthermore, we can see in Fig.~\ref{fig:fig1} that even if the present 
bound on the rare decay $\tau^-\to e^-\gamma$ improves by a factor of thousand
the cross section for $e^+e^-\rightarrow\tilde{\chi}^+_1\tilde{\chi}^-_1$ can 
change by a factor two when comparing the cross section for the
lepton flavour conserving (LFC) case $\cos2\theta_{13}=1$ with the one 
for which LFV is maximal ($\cos2\theta_{13}=0$).
We find that for $m_{\tilde\nu_3}=400$~GeV (solid line) and 
$m_{\tilde\nu_3}=600$~GeV (dashed line) a cancellation of one order of 
magnitude between chargino--sneutrino loop contributions and the 
neutralino--slepton loop contributions to BR$(\tau^-\to e^-\gamma)$ 
occurs in the (larger) right amplitude $A^R$, Eq.~(\ref{eq:raredecays}). 
The amplitude for the case where $m_{\tilde\nu_3}=600$~GeV
is a factor 6 larger (at $\cos2\theta_{13}=0$) compared to the case 
where $m_{\tilde\nu_3}=400$~GeV,
which explains the relative size of the corresponding branching ratios. 
For $m_{\tilde\nu_3}=900$~GeV (dotted line) no cancellation of the
various contributions to $A^R$ takes place. 
We note that the branching ratio BR$(\tau^-\to e^-e^+e^-)$ is 1--2 
orders of magnitude smaller than BR$(\tau^-\to e^-\gamma)$. 
We find that although the size of the cross section strongly depends
on the choice of the beam polarizations, the relative size of 
the cross section with and without LFV is almost independent of it.

\begin{figure}[t]
\setlength{\unitlength}{1mm}
\begin{center}
\begin{picture}(170,100)
\put(-60,-145){\mbox{\epsfig{figure=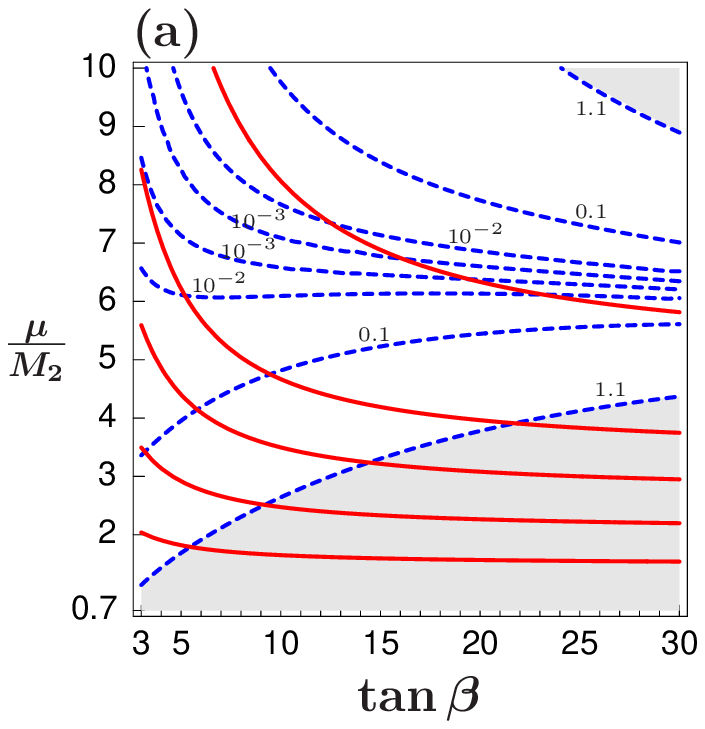,height=26cm,width=22cm}}}
\put(17,-145){\mbox{\epsfig{figure=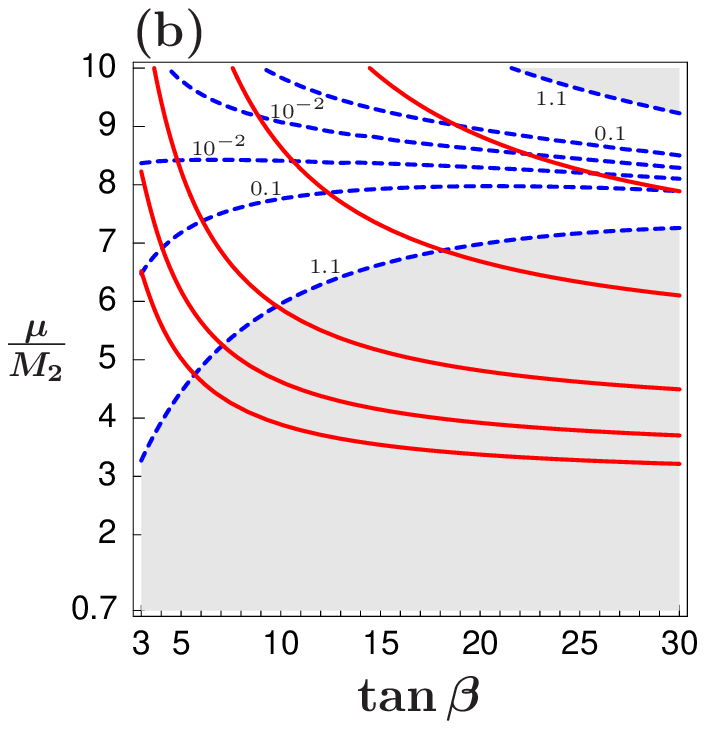,height=26.cm,width=22cm}}}
\end{picture}
\end{center}
\vskip-3cm
\caption{Contours of $10^7\cdot$BR($\tau^-\to e^-\gamma$) (dashed lines)
and $\sigma^{\rm LFV}_{11}/\sigma^{\rm LFC}_{11}$ (solid lines) in the
$\mu/M_2$--$\tan\beta$ plane. In (a) we have $m_{\tilde\nu_3}=400$~GeV with the 
contours $\sigma^{\rm LFV}_{11}/\sigma^{\rm LFC}_{11}=(1.5,1.7,1.8,1.85,1.9)$
(bottom-up), and in (b) we have  $m_{\tilde\nu_3}=900$~GeV with the 
contours $\sigma^{\rm LFV}_{11}/\sigma^{\rm LFC}_{11}=
(4,4.1,4.2,4.3,4.35)$ (bottom-up). The shaded areas in (a) and (b)
mark the regions excluded by the present experimental limit 
BR$(\tau^-\to e^-\gamma)<1.1\cdot 10^{-7}$.}
\label{fig:fig2}
\end{figure}

In Fig.~\ref{fig:fig2} we plot the contours of the branching ratio 
$10^7\cdot$BR($\tau^-\to e^-\gamma$) (dashed lines) and the contours of the ratio 
$\sigma^{\rm LFV}_{11}/\sigma^{\rm LFC}_{11}$ (solid lines) in the $\mu/M_2$--$\tan\beta$
plane, where we have used the abbreviations $\sigma^{\rm LFV}_{11}\equiv
\sigma(e^+e^-\rightarrow\tilde{\chi}^+_1\tilde{\chi}^-_1)$ for 
maximal LFV ($\cos2\theta_{13}=0$) and $\sigma^{\rm LFC}_{11}\equiv
\sigma(e^+e^-\rightarrow\tilde{\chi}^+_1\tilde{\chi}^-_1)$ for
the lepton flavour conserving case ($\cos2\theta_{13}=1$).
The other MSSM parameters are the same as in Fig.~\ref{fig:fig1}.
In Fig.~\ref{fig:fig2}a we show the result for $m_{\tilde\nu_3}=400$~GeV
where the contours of $\sigma^{\rm LFV}_{11}/\sigma^{\rm LFC}_{11}$
are 1.5, 1.7, 1.8, 1.85 and 1.9 for increasing $\mu/M_2$.
In Fig.~\ref{fig:fig2}b we have chosen $m_{\tilde\nu_3}=900$~GeV and
the contours for $\sigma^{\rm LFV}_{11}/\sigma^{\rm LFC}_{11}$
in this case are 4, 4.1, 4.2, 4.3 and 4.35 for increasing $\mu/M_2$.
As can be seen in Fig.~\ref{fig:fig2}a and b there is a
region in the $\mu/M_2$--$\tan\beta$ plane where the branching ratio
BR$(\tau^-\to e^-\gamma)$ is two to three orders of magnitude below
its present experimental bound and the values of the 
cross section $\sigma(e^+e^-\rightarrow\tilde{\chi}^+_1\tilde{\chi}^-_1)$ in the LFV case
can be about a factor 2 and 4 larger than in the LFC case.
In this region $\tilde\chi^+_1$ is gaugino--like and $\tan\beta$
can have any value in the range shown in Fig.~\ref{fig:fig2}.

\subsection{$\tilde\nu_e$--$\tilde\nu_\mu$ mixing case}

Now we consider the case of a non-vanishing $M^2_{L,12}$, putting
$M^2_{L,13}$ and $M^2_{L,23}$ to zero.
The size of $M^2_{L,12}$ is strongly restricted by
the experimental upper bounds on the 
LFV processes $\mu^-\to e^-\gamma$ and $\mu^-\to e^- e^+ e^-$
whose sensitivities are about four orders of magnitude larger
than those on LFV tau decays and will improve
substantially in the near future \cite{PSI}.
Similarly as in the previous subsection we take as our input parameters 
the sneutrino masses $m_{\tilde\nu_1}$, $m_{\tilde\nu_2}$, $m_{\tilde\nu_3}$
and the LFV mixing angle $\cos2\theta_{12}$ instead of the soft SUSY breaking parameters 
in the sneutrino mass matrix, Eq.~(\ref{eq:sneutrinomass}).

\begin{figure}[t]
\setlength{\unitlength}{1mm}
\begin{center}
\begin{picture}(150,50)
\put(-45,-145){\mbox{\epsfig{figure=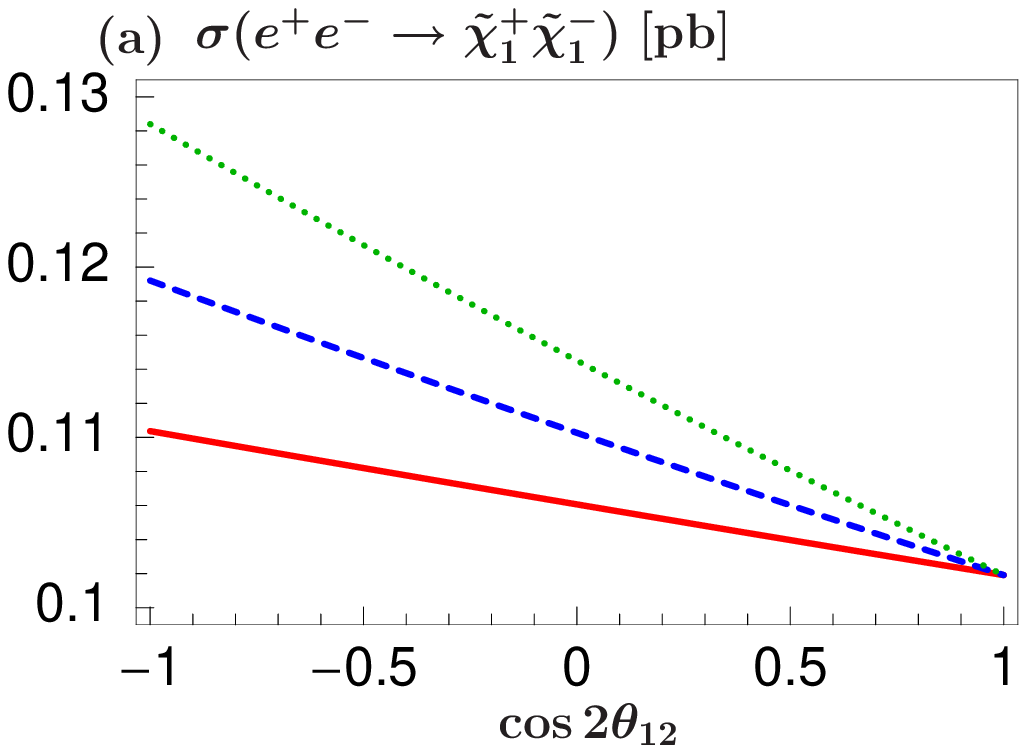,height=22.cm,width=14.4cm}}}
\put(30,-137){\mbox{\epsfig{figure=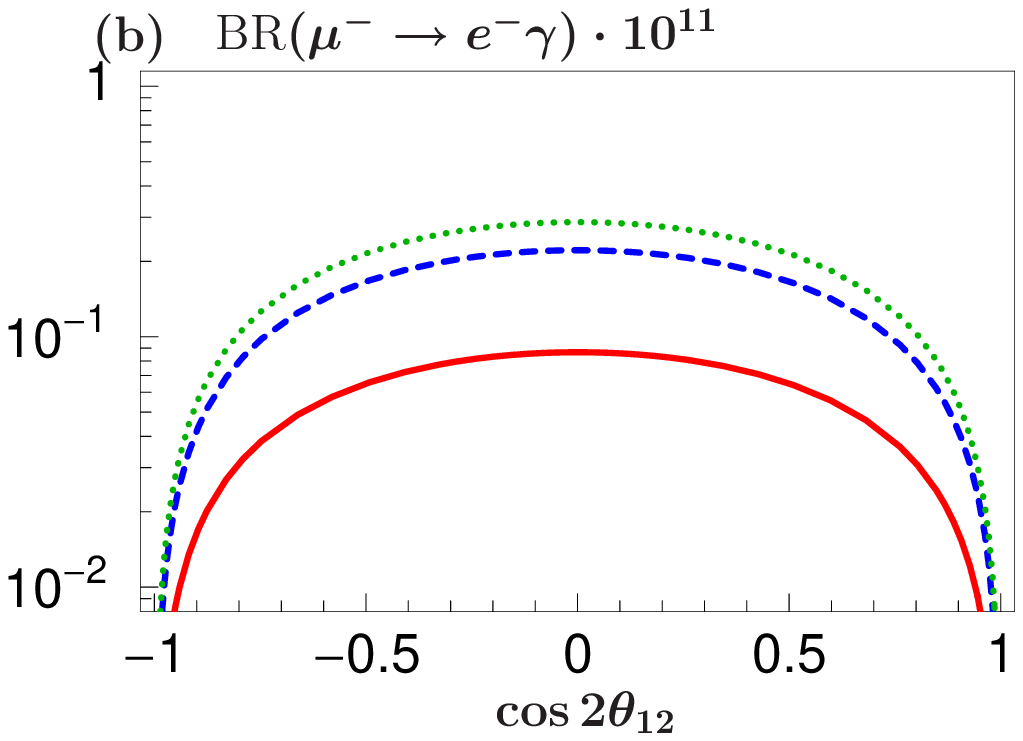,height=21.cm,width=14.4cm}}}
\end{picture}
\end{center}
\caption{(a) Cross section
$\sigma(e^+e^-\rightarrow\tilde{\chi}^+_1\tilde{\chi}^-_1)$ and
(b) branching ratio BR($\mu^-\to e^-\gamma$) as a function
of $\cos2\theta_{12}$. The three lines correspond to 
$m_{\tilde\nu_2}=305$~GeV (solid line),
310~GeV (dashed line) and 315~GeV (dotted line). 
The other parameters are as specified in the text.}
\label{fig:fig3}
\end{figure}

In Fig.~\ref{fig:fig3}a we show the $\cos2\theta_{12}$ dependence of the
cross section $\sigma(e^+e^-\rightarrow\tilde{\chi}^+_1\tilde{\chi}^-_1)$ 
for three values of $m_{\tilde\nu_2}=(305,310,315)$~GeV with $m_{\tilde\nu_1}=300$~GeV,
$m_{\tilde\nu_3}=500$~GeV, $\mu=1350$~GeV and the other parameters as 
defined in Fig.~\ref{fig:fig1}. 
The chargino masses are $m_{\chi_1}=237$~GeV and $m_{\chi_2}=1355$~GeV.
Fig.~\ref{fig:fig1}b shows the corresponding dependence of the branching ratio 
BR($\mu^-\to e^-\gamma$) for the same parameters.
The LFV mixing angle $\cos2\theta_{12}$ is not restricted
and can have any value in the whole range $[-1,1]$, where
for the values $\cos2\theta_{12}=-1,1$ lepton flavour is
conserved. Once $\cos2\theta_{12}\neq -1,1$
the sneutrinos $\tilde\nu_1$ and $\tilde\nu_2$ are mixtures
of the flavour states $\tilde\nu_e$ and $\tilde\nu_\mu$.
For $\cos2\theta_{12}=0$ they are a mixture containing an equal 
amount of $\tilde\nu_e$ and $\tilde\nu_\mu$, corresponding
to the case of maximal LFV. By comparing the cross sections
of the LFC case with $\cos2\theta_{12}=1$ and the case where 
LFV is maximal ($\cos2\theta_{12}=0$), we see from Fig.~\ref{fig:fig3}a
that the difference can be about 12\%.
For the three lines in Fig.~\ref{fig:fig1}b 
a cancellation of one order of magnitude between the chargino--sneutrino 
loop contributions and the neutralino--slepton loop
contributions to BR($\mu^-\to e^-\gamma$) occurs in the (larger) amplitude $A^R$,
Eq.~(\ref{eq:raredecays}).
We find that the branching ratio BR$(\mu^-\to e^-e^+e^-)$ is 1--2 
orders of magnitude below its present bound in this scenario. 

\section{Conclusions \label{sec:5}}

We have studied the production processes $e^+e^-\to\ti\chi^+_i\ti\chi^-_j$ 
in the MSSM including LFV terms. In the presence of non--vanishing 
LFV parameters in the sneutrino sector, the sneutrino contribution
to the chargino production process is different compared to
the case where these parameters are zero.
We have numerically studied the influence of LFV on the 
production cross section $\sigma(e^+e^-\to\ti\chi^+_1\ti\chi^-_1)$ 
and have found that this influence can be enormous.
We have demonstrated that $\sigma(e^+e^-\to\ti\chi^+_1\ti\chi^-_1)$ can 
change by a factor of 2 or more through non--vanishing
LFV parameters which are consistent at the same time with
the present limits on LFV rare lepton decays.
Moreover, we have pointed out that this statement holds
even in the case where the limit on BR($\tau^-\to e^-\gamma$) improves 
by a factor of thousand. In the effort of reconstructing the underlying
model parameters from measurements of chargino production cross sections,
it is therefore inescapable to take such a possibly sizeable effect
of LFV into account. This can done by measurements of lepton
flavour violating production and decay rates of SUSY particles 
at the linear collider, see e.g. 
\cite{Hisano:1998wn,Nomura:2000zb,Guchait:2001us,Porod:2002zy,Oshimo:2004wv,SneuLFV}. 
For example, a measurement of the event rates 
for the reaction $e^+e^-\to\tilde\nu\bar{\tilde\nu}\to\tau^+ e^-\tilde\chi^+_1\tilde\chi^-_1$
may allow one to determine the LFV mixing angle $\cos2\theta_{13}$ in the 
sneutrino sector \cite{Nomura:2000zb,SneuLFV}.

\section*{Acknowledgements}

We like to thank K. Hidaka, W. Majerotto and W. Porod
for useful and interesting discussions on this subject. 
We are grateful to A. Bartl for carefully reading the manuscript and
for his suggestions for improvement. 
This work is supported by the 'Fonds zur F\"orderung der
wissenschaftlichen Forschung' (FWF) of Austria, project. No. P18959-N16.
The authors acknowledge support from EU under the MRTN-CT-2006-035505
network programme.

\end{document}